\documentstyle[aps,twocolumn,eqsecnum]{revtex}
\begin{document}
\title{{\Large\bf Gaussian Wavepacket Dynamics: semiquantal and semiclassical
phase space formalism}}
\author{\bf Arjendu K. Pattanayak and William C. Schieve}
\address{Ilya Prigogine Center for Studies in Statistical Mechanics \&
Complex Systems,\\ Department of Physics, The University of Texas, Austin,
TX 78712}
\date{\normalsize 15th August, 1994}
\maketitle


\begin{abstract}
Gaussian wavepackets are a popular tool for semiclassical
analyses of classically chaotic systems. We demonstrate that they are
extremely powerful in the semiquantal analysis of such systems, too, where
their dynamics can be recast in an extended potential formulation.
We develop Gaussian semiquantal dynamics to provide a phase space formalism
and construct a propagator with desirable qualities.
We qualitatively evaluate the behaviour of these
semiquantal equations, and show that they reproduce the quantal behavior better
than the standard Gaussian semiclassical dynamics. We also show that these
semiclassical equations arise as truncations to semiquantal dynamics
non-self-consistent in $\hbar$.  This enables us to
introduce an extended semiclassical dynamics that retains the power of the
Hamiltonian phase space formulation. Finally, we show
how to obtain approximate eigenvalues and eigenfunctions in this formalism, and
demonstrate with an example that this works well even for a classically
strongly chaotic Hamiltonian.
\end{abstract}

\section{Introduction}
There has been a recent renewal of interest in the study of the
quantum-classical correspondence problem in many fields of physics,
termed 'postmodern quantum mechanics', in analogy with with the postmodern
movements in the humanities \cite{phystoday}. These studies are an attempt to
understand the characteristics of quantum systems in terms of the dynamics of
associated classical systems \cite{stev,prl}.
If the classical system is obtained in the canonical fashion by replacing
operators with $c$ numbers, the approach is the usual semiclassical
approach. The association here is straightforward only when the classical
system is integrable: the invariant torii of the classical phase space provide
the Einstein-Brillouin-Keller (EBK) quantization rules \cite{EBK}.
Matters are substantially more complicated when the classical system is
non-integrable, which is the more general case.
It is this $\hbar \to 0$ limit of the quantum mechanics of systems
where the classical ($\hbar \equiv 0$) dynamics are chaotic that is not
well-understood, and where the recent focus has largely centered
\cite{gutz,linda,eckhardt,leshouches}.
In particular, there has been work examining the
statistical properties of the spectrum of such systems \cite{linda} and the
remarkable calculation method known as the Gutzwiller Trace Formula, which says
that the eigenvalues of the quantum system arise as poles in a weighted sum
over the unstable periodic orbits of the classical system \cite{gutz}.

Chaos in classical mechanics is best observed in phase space, using
techniques such as Poincar\'e sections.
Hence, alongside the spectrum, there have been studies which attempt to
associate the complex structures of a classically chaotic phase space
with the eigenfunctions of the quantum system. These have included
the study of 'scars', which are the intensity peaks in an eigenfunction along
the unstable periodic orbits of the classical system \cite{oconnor,meredith}.
This idea also figures in the association of localized eigenstates with
structures such as invariant stable and unstable manifolds \cite{jensen} and
cantori (Birkoff-Gustavson quantization \cite{delos}).

Neither quantum mechanics nor semiclassical mechanics has a standard phase
space; all the analyses above must, perforce, carry out calculations in
Hilbert space and then project onto the classical phase space using 'symbols'
such as the Wigner Distribution Function or the Husimi Distribution
Function \cite{hillery}. This makes a direct interpretation of the results,
and the association of quantum properties with classical structures all that
more difficult.

There exists, however, another systematic approach to associating quantum
mechanics with a classical system. This may be motivated in many ways,
the simplest being the time-dependent variational principle (TDVP).
This approach starts with the variational restriction of Schr\"odinger's
equation to a subspace of the full Hilbert space. It can be shown that in this
approach, one always obtains a 'classical' phase space, where the dimensions
of the phase space correspond to the wavefunction parameters.
One can then derive and study the approximate (variationally guessed) dynamics
of the wavefunction in a straightforward manner. Since this technique starts
from the quantum Hamiltonian, with no reference to the classical system, and
can be shown to exist even for systems without a well-defined classical
dynamics, it has been termed semiquantal dynamics \cite{feng94,prl}.
In this paper we start with this semiquantal approach; we now sketch the
motivation for developing it.

Semiquantal dynamics studies a classical system (different from the canonical
one) to be able to understand the quantum system.
Since these dynamics derive as a variationally restricted form of the
Schr\"odinger equation, it follows that as the size of the Hilbert space
accessed is increased, the full quantum dynamics will be more closely
approximated by the classical system thus obtained.
[We point out that most numerical work involving 'exact' quantum mechanics
uses a superficially similar argument, wherein the wavefunction is expanded
in a basis set of finite size. There are fundamental differences between the
two approaches; in particular, even when the Hilbert space is inherently
non-compact, as in the systems we consider, basis set truncations are a
finite-dimensional restriction to a {\em compact} subset of the Hilbert space
whereas the semiquantal restriction is to a non-compact space.]
There is hence intrinsic value in understanding these approximate dynamics.
Further, as we will show, classical and semiclassical mechanics can be derived
as special limits or truncations of semiquantal mechanics.
This route thus provides a smooth transition, in principle,
between the two extremes of the classical limit and the fully quantal system.
The transition is made without, at any stage, sacrificing the access to
the considerable formal power of a classical phase space. It therefore
constitutes an alternative way to the
understanding of the quantum-classical correspondence question.
This provides the formal motivation; since semiclassical dynamics is contained
within semiquantal dynamics, the approach supplements the
semiclassical procedure without being detached from it.
Secondly, recognising the connection between
semiclassical and semiquantal dynamics enables us to propose a Hamiltonian
semiclassical dynamics, which combines the advantages of both approaches.
All the formalism of semiquantal dynamics carries over to this Hamiltonian
semiclassical dynamics; this is our practical motivation

Our discussion of semiquantal dynamics in this paper will be
through the use of coherent states\cite{klauder,sudarshan,glauber}.
These arise naturally in semiclassical limits and also in the limit
$N \to \infty$ where $N$ is the number of degrees of freedom of a
quantum system \cite{witten,yaffe}.
Coherent states come in various flavors, and can be constructed in any
quantum system if the dynamical group and the algebra are specified
\cite{skag,rmp,perel}. Their simplest version is in ordinary quantum mechanics
with a
Hamiltonian $\hat H = \hat p^2 /2 + V(\hat x)$; they are Gaussian wavepackets.
We work with this particularization; however,
most of the ideas are constructively generalizable.

Our presentation is organized as follows: in the next section we review the
derivation of the semiquantal dynamics of Gaussian wavepackets.
In Section 3, we derive the standard semiclassical dynamics of Gaussians,
following the work of Heller \cite{heller}.
We then show explicitly that those dynamics, derived as
consistent truncations in derivatives of the Hamiltonian, are not self-
consistent truncations in the moments of the wavepacket.
We thus propose an extended semiclassical dynamics 'midway' between these
semiclassical and semiquantal dynamics, which retains the Hamiltonian
formulation, and hence constitutes a Hamiltonian semiclassical dynamics.
We include a qualitative discussion of the comparative validity of these
three dynamics in various regions of a classically chaotic phase space, and
argue that the approaches developed in this paper improve upon the dynamics
derived by Heller. In Section 4, we then develop the formalism of the phase
space construction, including the explicit representation of arbitrary
wavefunctions and of the propagator.  Finally, we show how to obtain
approximate eigenvalues and eigenfunctions in this formalism, and
demonstrate with an example that this works well even for a classically
strongly chaotic Hamiltonian; we conclude with a short summary.

\section {Semiquantal dynamics of Gaussian wavepackets}

The basic idea behind the use of Gaussian wave packets relies on the following
\cite{Schro,heller,LJ}:
\begin{enumerate}
\item Since Gaussians are their own Fourier transform, they are well behaved in
phase space and their dynamics has an invariant meaning; unlike the
WKB method and the Van Vleck propagator\cite{gutz}, there are no singularities
that have to be dealt with by using 'matching formulas' etc.

\item Their dynamics are simply parametrized by the $c$ number variables that
specify the centroid (average variables) and spread (fluctuation variables).

\item They form an overcomplete set; an arbitrary wavefunction can be
expanded in terms of Gaussians (albeit an infinite number of them).
\end{enumerate}

This has lead to their use in methods known variously as the time-dependent
Gaussian variational approximation \cite{Jackiw,cooper,kovner},
the time-dependent Hartree-Fock-Boguliobov method or the
squeezed coherent state ansatz. The usual derivation considers the
time-dependent variational principle (TDVP) formulation \cite{dirac,KraSa},
wherein one posits the action
\begin{equation}\label{eq:tdvp}
\Gamma =\int dt < \Psi,t| i\hbar {\partial \over\partial t} - \hat H| \Psi, t>.
\end{equation}
The requirement that $\delta \Gamma = 0$ against independent variations of
$<\Psi,t|$, and $|\Psi,t>$ yields the Schr\"odinger equation and its
complex conjugate (note that a ray rotation, i.e. $| \Psi, t> \to
\exp (i\lambda (t)/\hbar)| \Psi,t>$ leaves the variational equations
unchanged).
The true solution is approximated by restricting
the choice of states to a subspace of the full Hilbert space
and finding the path along which $\delta \Gamma = 0$ within this subspace.

While the formal derivation of semiquantal dynamics relies on the TDVP,
we review here the alternative derivation of
semiquantal dynamics via the Ehrenfest theorem \cite{prl,diss};
we believe that this
is a simpler, somewhat more intuitive approach similar to formulations
in non-equilibrium statistical mechanics. It also helps us make the connection
with the standard semiclassical dynamics of Gaussian wavepackets in a
straightforward fashion. The dynamics obtained are the same as in the TDVP
approach. We work, for simplicity, with a one-dimensional problem.
The extension to multiple dimensions is presented in the appendix.

Consider, therefore, a particle of unit mass moving in a bounded potential with
a Hamiltonian $ \hat H = {\hat p ^2 \over 2} + V(\hat x)$ ( $V$ is
assumed analytic in $\hat x$ and ${\hat O}$ denotes operators).
Accordingly, the equations of motion for the centroid of a wavepacket are
\begin{eqnarray}
{d \over dt} < \hat x> & = & <\hat p> \\
{d \over dt} <\hat p>  & = & - <{\partial V (\hat x ) \over \partial x}>
\end{eqnarray}
where the $<>$ indicate expectation values. If $V$
is at most quadratic in $\hat x$ then the $<>$ factor;
\begin{equation}
<{\partial V (\hat x ) \over \partial x}> = {\partial V
\over \partial x} \bigg |_{<\hat x>} ,
\end{equation}
and the centroid therefore follows the classical trajectory.

The classical equations of motion do not hold for the centroid in the case of
nonlinear gradients, however, and to study such systems we expand in Taylor
series around $<\hat x>$ and $<\hat p>$. The expansions are of the form
\begin{equation}\label{eq:tay}
 <F(\hat u)> = {1 \over n!} <\Delta u^n> F^{(n)} \;\;  n \geq 0
\end{equation}
where $ F^{(n)} = {\partial ^n F / \partial u^n} |_{<\hat u>}, \Delta u =
\hat u - <\hat u>$ (we use the repeated index summation convention throughout
this section of the paper unless otherwise specified).
We note that if all the terms in the expansion are
retained, Eq. \ref{eq:tay} is an identity. Using this expansion and commutation
rules of operators, we can generate a series of moment equations.
In general these are infinite in number; an arbitrary distribution (wavepacket)
is completely specified only if all moments are known.  We have little
interest in the entire wave function and it would suffice to keep track only
of $<\hat x>$ and $<\hat p>$, thus creating an `Ehrenfest phase space'.
The convergence properties of the infinite hierarchy are not well understood;
it cannot be (or at least has not been) reduced in general .

A possible way \cite{andrews} of obtaining a finite set of
equations is to adopt a
Hartree-like  approach, requiring that all n-point functions be expressible
in terms of the one- and two- point functions.  This is equivalent to assuming
that the wavepacket is a normalized squeezed coherent state of the form
\begin{equation}
\Psi (q, t) = (2 \pi \mu)^{- {1 \over 4}}\exp \bigg [ i[ A(q - x)^2
 +  p\cdot(q-x)  ]\bigg ]
\end{equation}
and implies the relations
\begin{eqnarray}
<\Delta x^{2m}> & = & { (2m)! \mu^m \over m! 2^m}\;\; {\rm (no\ summation)} ,\\
<\Delta x^{2m+1}> & = & 0 ,\\
 4\mu <\Delta p^2> & = &  \hbar^2 + \alpha^2, \\
<\Delta x \Delta p + \Delta p \Delta x> & =  & \alpha ,
\end{eqnarray}
where
\begin{equation}
A = {1 \over 4\mu}( i + \alpha).
\end{equation}
[This assumption is the same as that of the TDVP: the wavepacket is
restricted to a given subspace. ]
This yields the equations
\begin{eqnarray}\label{eq:mu1}
{d x \over dt}  &= & p,\\
{d p \over dt}  &= & - {\mu ^m \over m! 2^{m}} V^{(2m+1)} (x), \; \; m= 0, 1,
 \ldots \\
{d \mu \over dt} &= & \alpha, \\
{d \alpha \over dt} &= &{ \hbar^2 + \alpha^2 \over 2 \mu}
-{\mu^m \over (m-1)!2^{m-2}}V^{(2m)}(x).\; \; m =1,2 \ldots \label{eq:muf}
\end{eqnarray}
The system is now reduced to the dynamics of $x, p,
\mu$ and $\alpha$ ( where we write $x,p$ for $<\hat x>,<\hat p>$)
and are exactly those derived from the action principle.
To see the equivalence, equate the variables
$x, p, \mu,$ and $\alpha$  with $q, p, \hbar G,$  and  $4 \hbar G \Pi$
respectively of the references \cite{Jackiw,cooper,tsue}.
[Note that we do not, for the moment, have
an equation for the phase of the wavepacket- this will be derived later.]

If we now introduce the change of variables $\mu = \rho^2,$ and
$\alpha = 2 \rho \Pi,$ Eqs. \ref{eq:mu1} through \ref{eq:muf} transform
to \cite{Rajagopal}
\begin{eqnarray}\label{eq:ext_dyn1}
{d x \over dt} &= & p ,\\
{d p\over dt}  &= &
 - {\rho ^{2m} \over m! 2^m} V^{(2m+1)} (x), \; \;  m= 0, 1, \ldots\\
{d \rho \over dt}  &= & \Pi, \\
{d \Pi\over dt} &= &{ \hbar^2  \over 4 \rho ^3}
-{\rho^{2m-1} \over (m-1)!2^{m-1}}V^{(2m)}(x), \; \;  m =1,2 \ldots
\label{eq:ext_dynf}
\end{eqnarray}

Remarkably, these new variables form an explicit gradient system, yielding an
{\em extended potential system} as our approximation to the Hilbert space.
The classical degrees of freedom are now the 'average' variables
$x,p$ and the 'fluctuation' variables $\rho, \Pi$, respectively; the
associated Hamiltonian is
\begin{eqnarray}
& H_{ext} & = {p^2 \over 2} + { \Pi ^2 \over 2 } + V_{ext}(x,\rho);\\
V_{ext} (x,\rho) = & V(x) &  + {\hbar ^2\over 8 \rho^2} +
{\rho^{2m} \over m!2^m }V^{(2m)}(x), \;\;  m =1,2 \ldots
\end{eqnarray}
where the subscript {\em ext} indicates the 'extended' potential
and Hamiltonian.

This formulation is very interesting and possibly quite powerful; the extended
potential provides a simple visualization of the geometry of the semiquantal
space. We may thus get a qualitative idea of the semiquantal
dynamics before proceeding to detailed analysis.

We note here, for instance, that:
\begin{enumerate}
\item Both the fluctuation and average variables are treated on the same
footing
and the phase space is dimensionally consistent: $\rho$ has the dimensions
of length and $\Pi$ that of momentum,
\item The value of $H_{ext}$ is $<\hat H>$ under this approximation,
and is conserved,
\item $V_{ext}$ has an infinite barrier at $\rho = 0$,
arising, as appropriate, from the momentum squeezing. Hence,
'quantum fluctuations' can not be zero except in the limit $\hbar \to 0$: as
stated, there are no caustics in this representation.
\item $V_{ext}$ is symmetric in $\rho$ corresponding to a choice of sign in
$\rho =\sqrt \mu = \sqrt {<X^2>} $; the infinite barrier renders the choice one
of convenience and of no physical significance.
\end{enumerate}

It is now straightforward to introduce a Poisson bracket on this space:
\begin{equation}\label{eq:Poisson}
 \{ f, g \} = {\partial f \over \partial x} {\partial g \over \partial p}
+ {\partial f \over \partial \rho} {\partial g \over \partial \Pi}
- {\partial f \over \partial p} {\partial g \over \partial x}
-{\partial f \over \partial \Pi} {\partial g \over \partial \rho}.
\end{equation}
This bracket renders the dynamics into the standard form
\begin{equation}
\dot y = \{y ,H_{ext} \}
\end{equation}
where $y$ represents the phase space variables $x,p,\rho,\Pi$.
The rate of change of an arbitrary function $f(y)$ is
\begin{equation}\label{eq:poiss}
{ df \over dt} = \{ f, H_{ext}\}.
\end{equation}

This Hamiltonian picture for the evolution of the wavepacket parameters
is not quite the complete  one since, as before, a ray rotation
leaves the expectation values unchanged. To determine the overall phase of the
wavepacket, therefore, consider the Schr\"odinger equation itself
\begin{equation} \label{eq:Schro}
i\hbar {\partial \over \partial t}| \Phi, t> = \hat H |\Phi, t>
\end{equation}
where
\begin{equation} \label {eq:rot}
|\Phi, t> \equiv exp(i\lambda (t)/\hbar) | \Psi, t>
\end{equation}
If we substitute eq. \ref{eq:rot} into eq. \ref{eq:Schro}, and take the inner
product with $|\Psi, t>$, we get for $\lambda (t)$ the differential equation
\begin{equation}\label{eq:phase}
{ d \lambda \over dt} = < \Psi,t| i\hbar {\partial \over \partial t}
- \hat H| \Psi, t>.
\end{equation}
The right-hand-side of the above equation is the Lagrangian corresponding to
the extended Hamiltonian (see \ref{eq:tdvp}). The phase is then just the action
function $\Gamma$ (divided by $\hbar$) and its explicit time-derivative is:
\begin{equation}
\dot \lambda = {\dot \rho \Pi - \dot \Pi \rho \over 2}+ p\dot x - H_{ext}.
\label{eq:act1}
\end{equation}

It is natural to separate the phase into two parts
$\lambda = \lambda _D + \lambda _G$ with the first part
\begin{eqnarray}
\lambda _D & = & - \int_0^t d\tau\;<\hat H>\nonumber \\
           & = & -t \;H_{ext}
\end{eqnarray}
corresponding to the dynamical phase (i.e. a measure of the time of evolution).
The second part is the geometrical phase:
\begin{eqnarray}
\lambda _G & = & \int_0^t d \tau\; < i\hbar {\partial \over \partial \tau}>
\nonumber\\
  & = & \int_0^t d \tau \;({\dot \rho \Pi - \dot \Pi \rho \over 2}+ p\dot x ).
\label{eq:lg1}
\end{eqnarray}
For the case of cyclic evolution of the wavepacket this corresponds to the
'Berry's phase' \cite{wilczek} of the evolution. It depends only on the
geometry of the evolutionary path in phase space since it can be written in the
canonical form
\begin{equation} \label{eq:lg}
\lambda _G (C) = \oint _C {\bf P \cdot d Q},
\end{equation}
where ${\bf P}\equiv (p,\Pi)$ and ${\bf Q}\equiv (x,\rho)$. Note that the
integrand of \ref{eq:lg1} differs from that of \ref{eq:lg} by a total
derivative which
vanishes for the closed line integral. The expression for $\lambda _G (C)$ has
this simple form because of the Hamiltonian character of semiquantal dynamics.
We will use this later for the construction of eigenvalues and eigenfunctions.

The equation for the phase, along with the Hamiltonian equations of motion for
the evolution of the wavepacket parameters constitute the semiquantal dynamics
of the wavepacket. This lies on a space ${\cal R}(2N) \times S (1)$; for the
case just considered, $N = 2$, the result is completely general, however.
Finally, we note that the overall normalization factor for the
Gaussian wavefunction is a trivial function of $\rho$ and does not consitute
an independent variable. The normalization is maintained by the evolution.

To summarize, the semiquantal restriction of the dynamics to Gaussian
wavepackets has taken us from a Hilbert space to a extended phase space and
the dynamics are described on this phase space by a generalized Poisson
bracket.
An arbitrary operator of the Hilbert space may be associated with a function
(called its 'symbol') through
\begin{equation}\label{eq:sym}
O (y) =  <y|\hat O|y>
\end{equation}
where $y =(x,p,\rho,\Pi)$  and the $|y>$ are normalized Gaussians serving as
the basis set for the association. The dynamics of this symbol are determined
by Eq. \ref{eq:poiss}: the Heisenberg equations of motion are now represented
by Poisson bracket equations. These are not the classical equations of motion,
however, since the generator of time translations is not the classical
Hamiltonian, but $H_{ext}$. We emphasize that for the expectation values of
operators, the space ${\cal R}(2N)$ (the approximation to the projective
Hilbert space) suffices for a complete description of the dynamics. The
dynamics of the wavefunction, however, also require the specification of the
phase; this is what lives on ${\cal R}(2N) \times S (1)$.

As with any variational principle, these equations are approximations to the
same extent as the restriction of the variational space. Similar equations
hold for the full space: the actual specification of an appropriate variational
wavefunction may not, of course, be possible.

\section{Comparison with semiclassical dynamics}

In this section, we first derive the standard semiclassical dynamics of
Gaussian wavepackets, following
Heller \cite{heller} (we hence term these dynamics, with apologies to others,
Heller's semiclassical dynamics) . We demonstrate that though these equations
arise as consistents truncations to semiquantal dynamics in
derivatives of the Hamiltonian, they are not self-consistent truncations
in $\hbar$. We thus are able to introduce an extended semiclassical dynamics;
this restores the consistency of the above truncation, and is a Hamiltonian
semiclassical dynamics. We finally compare the three approaches - semiquantal
dynamics, Heller's semiclassical dynamics and extended semiclassical dynamics.

We make a preliminary note that any truncation to the sums in the
equations \ref{eq:ext_dyn1} through \ref{eq:ext_dynf} at a specified order
in $m$, say the $nth$,
retains the Hamiltonian structure. Its properties (as developed in the next
two sections) thus transfer over to the
truncated systems. The only detail that one has to be careful about is the
evaluation of the symbols for operators. This consideration and the
interpretation of the truncations is simplified by the standard
semiclassical ansatz that
\begin{equation}
 <\hat x - <\hat x>>^2 = \rho ^2 =  {\cal O}(\hbar).
\end{equation}
This implies, therefore, that
\begin{equation}
<\hat p - <\hat p>>^2 = {\cal O}(\hbar) = \Pi ^2.
\end{equation}
The truncations are then to the same order (the $nth$) in $\hbar$.
The calculation of the symbols is straightforward - one
proceeds with the computation as if for the full semiquantal dynamics
and then retains only terms to the specified order in $\hbar$.

\subsection{Heller's semiclassical dynamics}

The semiclassical derivation of the equations of motion for Gaussian
wavepackets, following Heller \cite{heller} makes an expansion around the
wavepacket centroid, truncated to second order in derivatives of the
Hamiltonian. We make the same simplifications as before;
the Hamiltonian is ${\hat p^2 \over 2 } + V(\hat x) $
and we use a one-dimensional wavepacket
\begin{equation}
\Psi (q, t) = \exp\bigg [{ i \over \hbar }[ (q - x)A
(q - x)   +  p\cdot(q-x) + s)]\bigg ].
\end{equation}
The physical meaning of the parameters $p(t),x(t)$ are, as before
$<\hat x(t)>  =  x(t), <\hat p(t)>  = p(t)$
and $A$ and $s$ are, in general, complex.

The fundamental approximation is
a truncation of the Taylor series expansion of the Hamiltonian as
\begin{equation}
H \approx -{\hbar ^2 \over 2}{\partial ^2 \over \partial q^2}
+ V^{(0)} + V^{(1)} (q - x) + {1 \over 2}V^{(2)} (q - x)^2,
\end{equation}
where $V^{(n)}$ is the $nth$ derivative of $V(\hat x)$ evaluated at $x$,
as before. This is exact for quadratic potentials, of course. If we insert
this, and the wavefunction above, into Schr\"odinger's
equation, equate coefficients of $(q - x)$ , and make the identification
$ \dot x = p $, we recover classical equations of motion for the centroid:
\begin{eqnarray}\label{eq:class1}
\dot x & = & p ,\\
\dot p & = & - V^{(1)}\label{eq:class2}
\end{eqnarray}
and obtain, for the wavepacket parameters
\begin{eqnarray}\label{eq:A}
\dot A  &=&  - 2 A^2  - {V^{(2)}\over 2},\\
\dot s  &=&   i\hbar A -  V(x) - {p^2 \over 2 } + p\dot x\label{eq:s}.
\end{eqnarray}

At this point, using our knowledge of the derivation of semiquantal Gaussian
dynamics, we make the transformation
\begin{eqnarray}
s &=& s_r + is_i\\
A & =& a_r + ia_i\nonumber\\
  & = & {\Pi \over 2\hbar \rho} + i {\hbar \over 4 \rho ^2}\label{eq:a2pi}
\end{eqnarray}
These then provide us the equations
\begin{eqnarray}
\dot s_r & = & -{\hbar ^2 \over 4 \rho ^2} - V(x) - {p^2\over 2}  + p\dot x\\
\dot s_i & = & {\Pi \over 2\rho}\\
\dot \rho & = & \Pi\label{eq:Hell1}\\
\dot \Pi & = & {\hbar ^2 \over 4 \rho ^3} - \rho V^{(2)}
\end{eqnarray}
along with
\begin{eqnarray}
\dot x & = & p ,\\
\dot p & = & - V^{(1)}\label{eq:Hell4}
\end{eqnarray}
as the equations of motion for the wavepacket.

It is straightforward to show that $s_i$ is
the time-dependent normalization term \cite{heller}. The solution for it is
$s_i(t) = {1 \over 2}ln(\rho)$ modulo a constant; we ignore it hereafter,
except
to note that it is the same as the semiquantal case.
The equation for $s_r$ can be rewritten, using the relationships
\begin{eqnarray}
\rho \dot \Pi &=& {\hbar ^2 \over 4 \rho ^2}- \rho ^2 V^{(2)},\\
\dot\rho\Pi &=& \Pi ^2,
\end{eqnarray}
as
\begin{equation}
\dot s_r  =  p\dot x + { \dot\rho\Pi - \dot\Pi \rho \over 2} - E
\end{equation}
where
\begin{equation}
E =  {p^2 \over 2}+ {\Pi ^2\over 2} + V(x) + {\hbar ^2 \over 8 \rho ^2}
+ {\rho ^2 \over 2}V^{(2)}.
\end{equation}
Note that this equation for the phase is the same as Eq. \ref{eq:act1} with
$E$ replacing $H_{ext}$ and that
\begin{equation}
E = H_{ext} \bigg |_{m=1}.
\end{equation}

This system of equations has been used with great success to study
semiclassical
dynamics by Heller and co-workers \cite{heller} and Littlejohn \cite{LJ},
amongst others. In particular,
it has been used to construct the semiclassical spectra of various sytems, by a
method we shall sketch later. Littlejohn has argued \cite{lj} that the
spectrum thus computed for integrable classical systems is exactly the EBK
spectrum. For non-integrable systems, it provides an alternative to basis-set
quantization and the Trace Formula. The interesting question of the connection
between these methods remains an open one.

\subsection{Extended Semiclassical dynamics}

How does this system compare with the semiquantal system we had derived
earlier ? It is a simple excercise to show that
equations \ref{eq:Hell1} through \ref{eq:Hell4} above are truncations to $m=1$
of the semiquantal equations \ref{eq:ext_dyn1} through \ref{eq:ext_dynf},
{\em except} for a missing term ${\rho ^2 \over 2}V^{(3)}$ in Eq.
\ref{eq:Hell4}
for $\dot p$. The phase, as mentioned, is also a truncation of the semiquantal
phase to $m=1$. The structural correspondence between the two systems is
then exact, except that Eq. \ref{eq:Hell4} is truncated to $m=0$ compared to
the others .

We have thus demonstrated that the standard equations for
the semiclassical dynamics of wavepackets correspond to truncations
of the semiquantal equations derived earlier.  These truncations are to
different values in $m$; there is hence no extended Hamiltonian for these
equations and the formalism developed below does not apply.
However, it is now easy to construct a semiclassical Hamiltonian.

We restore consistency in the truncation by adding back the 'extra' term
${\rho ^2 \over 2} V^{(3)}$, in the equation for $\dot p$.
This term arises naturally in the expansion around the centroid, but was
neglected by the truncation to the second derivative. Hence, arguments similar
to the above, truncated not in derivatives of the Hamiltonian, but moments of
the Gaussian, give us an extended semiclassical Hamiltonian
\begin{equation}
H_{sc} =  {p^2 \over 2}+ {\Pi ^2\over 2} + V(x) + {\hbar ^2 \over 8 \rho ^2}
+ {\rho ^2 \over 2}V^{(2)}.
\label{eq:Hsc}
\end{equation}
This Hamiltonian is a truncation of the semiquantal Hamiltonian to $m=1$ and
by the arguments above, to ${\cal O}(\hbar)$. This is what we call extended
semiclassical dynamics - it lies 'midway' between the standard semiclassical
dynamics and semiquantal dynamics.

It is now simple to show that these systems contain within themselves the
classical limit. If for each system, we take the physical limit $\rho, \Pi \to
0$ along with $\hbar \to 0$,  we get a Gaussian that corresponds in the limit
to a classical point particle, and the dynamics reduce to that of the classical
Hamiltonian. Equivalently, we could make the same ${\cal O}(\hbar)$ ansatz as
before, and take $\hbar \to 0$ to recover the classical equations.

We now compare the three approaches: semiquantal dynamics,
Heller's dynamics and extended semiclassical dynamics and discuss their
validity and utility.

\subsection{Similarity of the three aproaches}

We first wish to emphasize that all three approaches are similar in
principle. The simplest way of seeing this is to consider their derivation in
the Heisenberg picture, which we now sketch.
This is a straightforward excercise that we do not belabor here.
First rewrite the quantum Hamiltonian as follows:
\begin{eqnarray}
\hat H & = & {\hat p^2 \over 2 } + V(\hat x)\\
       & = & {p^2 \over 2 } + p\Delta p + \Delta p^2
        +  V(x) + \sum_{n=1}{1 \over n! } \Delta x^n V^{(n)}
\end{eqnarray}
where $ u = <\hat u> , \Delta u = \hat u - u$ and $  V^{(n)} =
 V^{(n)} = {\partial ^n V / \partial x^n} |_{<\hat x>} $ as usual.
Now use this (still exact) form of the Hamiltonian to derive
equations of motion for the Heisenberg operators $\hat x, \hat p,
\Delta x ^2 $ and $ \Delta x\Delta p + \Delta p \Delta x$ (or any other
operator for that matter). Having obtained these operator equations
turn them into $c$-number equations by {\em imposing} a Gaussian ansatz
and evaluating the expectation values of these equations in that ansatz.
These equations are now the semiquantal equations derived earlier [the
equations for the higher moments are now redundant, as before, and may
be safely ignored].
If, in the above expansion of the potential, we truncate at $n=2$, and
proceed as outlined, again imposing the Gaussian ansatz,
we obtain Heller's equations.
A truncation at $n=3$ yields the extended semiclassical
dynamics.  All three approaches are hence identical in nature.
In particular, we note that all three are {\em local} expansions around
the centroid. The global Hamiltonians $H_{ext}$ and $H_{sc}$ are effective
Hamiltonians that the wavepacket sees, and are obtained by no more a global
approximation than for the standard semiclassical case.
The only point to consider is that the Gaussian ansatz seems 'natural' in the
quadratic  approximation; however, as far as the real system is concerned, it
is
just as valid (or invalid) a constraint for all the approaches.

We also note that both the semiquantal and extended
semiclassical dynamics can also be derived as examples of 'symbol' dynamics
\cite{Voros}. In this approach, we project out a $c$-number symbol
of the Hamiltonian (or its local quadratic approximation, respectively)
in appropriate states (the squeezed coherent states), and treat
this as a classical Hamiltonian, with an eye to understanding the
quantum behavior. Heller's semiclassical dynamics are similar, but not
Hamiltonian because the projection onto the squeezed coherent states is done
after the dynamics are derived, instead of before, leading to an
inconsistency in the truncation.

For the semiclassical approaches, a time-scale of validity
\cite{hage} for the ansatz may be estimated and is normally argued to be
${\cal O}(ln (\hbar ^{-1}))$, the 'log time'. Recent
work \cite{phystoday}, however, argues that this may be unduly pessimistic.
Further, as we demonstrate below, many {\em qualitative} features of quantum
behavior can certatinly be reproduced by these dynamics.

\subsection{Qualitative behavior in model potentials}

We now discuss how the qualitative behavior of these approaches conforms to
intuition. We examine these dynamics in a few simple model Hamiltonian
corresponding to local neighborhoods of a classically chaotic phase space. We
start with the corresponding quantal Hamiltonian and derive from it the various
dynamics. We then consider the behavior of wavepackets that start centered or
slightly displaced from the origin in these model potentials.
We will not go into the details; they can be easily worked out from the
results of the last two sections.

We first consider the more forgiving case of elliptic regions (equivalently,
in general, an anharmonic well) where such a Gaussian ansatz might be
expected to work. Our model Hamiltonian is
\begin{equation}
\hat H = {\hat p^2 \over 2 } +  \hat x^{2n}
\end{equation}
[where we consider $ n > 1$ since all three dynamics are identical and
exact in quadratic
potentials].  The potential is bounded in $\hat x$ and its highest
non-vanishing
even derivative hence is positive. For the semiquantal dynamics, the form of
the semiquantal Hamiltonian then guarantees that $\rho$ always remains
bounded as well; this is physically reassuring.  However, this fails for the
two semiclassical systems. In particular, for the simple $ \hat x^4$
potential, the dynamics for $\rho$ for both semiclassical systems are
unphysically unstable.  This is easiest seen by considering the wavepacket
centred at the origin ($x(t) \equiv x(0) = 0$) where $V^{(2)}$ identically
vanishes. The situation improves when the wavepacket is displaced from the
origin though the dynamics are still asymptotically unstable with the extended
semiclassical system performing better than Heller's dynamics due
to the coupling between the average and fluctuation variables.

We now look at hyperbolic regions. If the model potential is unbounded, the
semiquantal approach {\em will} always give unstable solutions. This is
not a breakdown of the method, but only as appropriate: an unbounded potential
{\em must} give a spreading wave-packet. However, note that
for the unbounded Hamiltonian
\begin{equation}
\hat H = {\hat p^2 \over 2 } + a\hat x^2 - \hat x^4
\end{equation}
($ a >0$) both Heller's dynamics and the extended semiclassical dynamics,
will be deceived into remaining stable with a centered wavepacket
except in the singular limit $a \equiv 0$.
This persist even for small displacements with finite $a$ for Heller's
dynamics.
For the extended semiclassical system the coupling of the average and
fluctuation dynamics kicks in to drive the wavepacket away from the
origin and give a spreading wave-packet.

We turn to the more appropriate case of hyperbolicity embedded in a bounded
potential, with a model Hamiltonian hence, say, of the form
\begin{equation}
\hat H = {\hat p^2 \over 2 } - \hat x^2 + \hat x^4.
\end{equation}
If we once again start centered at the origin, the semiquantal approach
provides for a growing wavepacket that is ultimately reigned in by the quartic
term, as is the appropriate physical behavior.
Both semiclassical systems will again be deceived - in this case into
rapidly spreading without a bound. With small displacements, the semiquantal
behavior is unaltered. Both semiclassical dynamics improve marginally and can
be stabilized, but in general remain unstable.

These discussions demonstrate that in these model potentials,
the semiquantal approach improves substantially upon Heller's semiclassical
Gaussian ansatz, while the extended semiclassical approach does so at least
marginally. The arguments above are not the whole story, of course, since the
wavepacket will in general be moving through these regions of phase space
instead of being statically centred there, but capture the essential features
of the expected behavior. It should be clear that the coupling between
the fluctuation and average variables is in general not simple to analyze.
This is a physically valid coupling and has its advantages, as just argued,
but may lead to behavior which has to be interpreted carefully.

\subsection{Interpretation}

In a recent paper \cite{prl}, it was demonstrated that the semiquantal dynamics
of the double-well system are chaotic, even though the classical and quantum
problem are understood to be regular. This is to be viewed largely as a
formal result: the transition between classical and quantum dynamics is not
quite as straightforward as is usually understood; quantal effects
do {\em not} always suppress chaotic behavior.
Further, an examination of the range of parameters and initial
conditions within which chaos appears indicates the following realistic
interpretation. The dynamics of the wavepacket both deep within either well,
and at energies such that the system looks essentially like an anharmonic well,
are regular. It is in the transition regime, at energies close to the classical
unstable fixed point and separatrix, that the dynamics are complicated.
It is the interplay between classically unstable behaviour and quantal
tunneling effects that are reflected in the observed chaotic tunneling. By the
methods outlined below, this would give rise to a rather
complicated spectrum in this region - a wholly appropriate result.

We reiterate that uncoupling the variables, thus yielding Heller's
semiclassical dynamics, would do worse in reflecting
the quantal behavior in this system, as outlined above.
Irrespective, it is obvious that this sort of behavior needs to be better
understood. The discussion in the rest of the paper leads us towards
doing so: matters do not end at having observed chaos or lack thereof in these
classical-like dynamics. This structure is reflected in the spectrum and
other pure quantal properties.

Considering all these factors, therefore, we argue for the use of semiquantal
and extended semiclassical dynamics in exactly the same situations as the
Heller's semiclassical Gaussian dynamics.
Wavepacket dynamics must ultimately break down in the presence of chaotic
classical dynamics; however, the two Hamiltonian approaches should better
reproduce the quantal behavior and thus improve upon the
already remarkable results of Heller's semiclassical dynamics.
The spirit of these approaches,  we believe, is truly semiclassical; the
Hamiltonian formulations make them particularly powerful.
This is developed in the following.

\section{Phase-space properties of the representation}

To understand formal power and properties of the Hamiltonian formulations
constructed above, we briefly discuss the properties of coherent states
using the standard notation, simplifying our discussion considerably;
we relate the notations later.

If we consider the non-Hermitian operator
\begin{equation}
\hat a = {1\over \sqrt 2} ( \hat x + i\hat p)
\end{equation}
with
\begin{equation}
[\hat a, \hat a^{\dagger}] = 1
\end{equation}
we can construct the states $|\alpha>$ defined as the  eigenstates of $\hat a$
\begin{equation}
\hat a |\alpha> = \alpha |\alpha>.
\end{equation}
These states are minimum uncertainty Gaussians, known as coherent states.
Due to the non-hermiticity of
$\hat a$, they are not orthonormal, nor are their eigenvalues real.
The properties of coherent states \cite{skag,gard} are well known; we mention
two we use later.
Coherent states have the following representation in
terms of the number states, the energy eigenstates of the harmonic oscillator
(in this case with $m = \hbar =\omega =1$):
\begin{equation}\label{eq:exp n}
|\alpha> = \sum e^{( -{1\over 2} |\alpha|^2) } {\alpha ^n \over \sqrt {n!}}|
n>.
\end{equation}
and they form an overcomplete basis set.

The coherent states do not include all Gaussians: we parametrize the space of
all Gaussians (for a one-dimensional system) using four real parameters
$x,p,\rho,\Pi$ whereas the coherent states use just two real parameters,
\begin{equation}
\alpha =  \alpha _1 + i \alpha _2.\\
\end{equation}
To understand the extra parameters, we consider squeezed coherent states.
We follow Yuen \cite{yuen} in introducing an operator $\hat b$
\begin{equation}\label{eq:a2b}
\hat b \equiv \mu \hat a + \nu \hat a^{\dagger}
\end{equation}
with the $c$ numbers $\mu, \nu$ satisfying
\begin{equation}\label{eq:munu}
|\mu|^2 - |\nu|^2 = 1.
\end{equation}
The real and imaginary parts of this equation imply that
$\mu$ and $\nu$ correspond to two independent real parameters.
It can be shown that
\begin{equation}
[\hat b, \hat b^{\dagger}] = 1,
\end{equation}
and therefore that the transformation \ref{eq:a2b} is a linear canonical
transformation.  This provides $\hat b$ with properties identical to
those of $\hat a$. In
particular, there exist a set of eigenstates of $\hat b$ ( which we denote
$ | \beta>_q \equiv | \alpha _1, \alpha _2, \mu, \nu> $ ) equivalent to
the states $|\alpha>$: these are the squeezed coherent
states. These states include the coherent states as a particular case,
where $ \mu =1$ and $\nu = 0$, and in general include all Gaussian
wavefunctions. The independent parameters $\alpha _1, \alpha _2, \mu, \nu$,
provide us with our four-parameter representation.

If we consider the self-adjoint operators $\hat a _1,\hat a _2$ defined through
$\hat a = \hat a_1 + i \hat a_2$ (corresponding to the quadrature components)
we see that
\begin{equation}
< \Delta a^2_1> = <\Delta a^2_2> = {1\over 4}.
\end{equation}
($\Delta m  = \hat m - <\hat m>$ ) where
all expectation values are taken with respect to the states $|\alpha>$. Since
$[\hat a_1,\hat a_2] = {i\over 2}$, the uncertainty principle constraint on
$\hat a_1, \hat a_2$ is
\begin{equation}
<\Delta a^2_1><\Delta a^2_2> \;\geq {1 \over 16}.
\end{equation}
The minimum uncertainty states $|\alpha>$ have equal uncertainty in the two
quadrature components; their correlation is zero:
\begin{equation}
< \Delta a_1 \Delta a_2> =0.
\end{equation}
If we evaluate the same quantities in the states $| \beta >_q$, we get
\begin{eqnarray} \label{eq:r1}
< \Delta a^2_1>_q &=& {1 \over 4} | \mu - \nu|^2,\\
<\Delta a^2_2>_q &=& {1 \over 4} | \mu + \nu|^2,\\
< \Delta a_1 \Delta a_2>_q &=&{1 \over 4} i ( \mu ^*\nu - \nu ^*\mu +1).
\label{eq:r3}
\end{eqnarray}
The squeezed states, therefore, 'squeeze' one quadrature component at the
expense of the other, and in general have correlations between the two
components. However, there is a linear canonical transformation of variables
$ (\hat a \to \hat b) $ such that the new quadrature components $ \hat b_1,
\hat b_2$ satisfy the minimum uncertainty equation in these states.
We use here Eqs. \ref{eq:r1} through \ref{eq:r3} to make the connection
with our previous notation. $x$ and $p$ are the same as above, and
\begin{eqnarray}
\rho ^2 &=& {1 \over 2} | \mu - \nu|^2,\\
\rho \Pi &=& {1 \over 2} i ( \mu ^*\nu - \nu ^*\mu ).
\end{eqnarray}
We now use these properties for the representation of arbitrary wavefunctions.

We have so far considered
the dynamics of a wavepacket that is constrained to remain Gaussian:
this yields trajectories in a generalized phase space.
However, it is also possible to construct propagators for arbitrary
wavefunctions. In this case we have to consider functions on this
phase space. We use the fact that
it is possible to associate with every operator a symbol as in Eq.
\ref{eq:sym}.
and choose $\hat O$ to be the density matrix (or projection operator)
$\hat P = | \Psi><\Psi|$.  We now associate with every
wavefunction (upto an arbitrary global phase) a real positive function
\begin{equation} \label{eq: rep}
{\cal P} (y) = <y| \Psi><\Psi|y>.
\end{equation}
This function is an extended Q-representation
(or extended Husimi function) : we now demonstrate this,
and touch upon its relevant properties.

With the coherent states as defined above,
the Q-function corresponding to a projection operator
$\hat P$ is defined as \cite{gard}
\begin{equation}\label{eq:Q}
Q(\alpha, \alpha ^*) \equiv {1 \over \pi}< \alpha |\hat P| \alpha >.
\end{equation}
The Husimi function is defined in almost the same way,
upto the normalization of the coherent states. Since the normalization of
the coherent states is well defined, we use the terms interchangeably.

The Q-function is essentially a probability distribution function, with useful
properties for the computation of expectation values of operators,
etc. \cite{gard}.  We do not use it directly, however.
since the dynamics we have derived do not leave
coherent states invariant. Hence, the dynamics in the Q-representation are
awkward and the Hamiltonian formulation cannot be used to advantage.
However, the space of squeezed coherent states is invariant under the derived
dynamics. We take advantage of this by defining, as an intermediate
tool, an extended Q-function over the space of squeezed coherent states.
The semiquantal dynamics of this can be easily handled - we detail this below.
The Q-function (and the projection operator itself) can also be extracted
from the extended Q-function in a straightforward fashion -
we first take care of this.

We define the extended Q-function $Q_{ext}(\alpha,\mu,\nu )$ as:
\begin{equation}
Q_{ext}(\alpha, \mu ,\nu) = < \alpha, \mu , \nu |\hat P | \alpha, \mu,\nu>
\end{equation}
where we absorb normalization constants into the definition of
the states. We have also suppressed the dependence of $Q_{ext}$ on the complex
conjugates of $\alpha, \mu , \nu$.
This function is precisely the representation we have derived
through semiquantal dynamics: $ Q_{ext} \equiv \cal P$.
It is defined over a four parameter space $\alpha _1, \alpha _2, \mu, \nu
$ and has the obvious and convenient property that a slice \cite{fn1} of it,
taken at $\mu =1, \nu =0$, reduces to the Q-function:
\begin{equation}
Q_{ext}(\alpha, \mu, \nu) \bigg |_{\mu =1,\nu =0}= Q(\alpha, \alpha ^*).
\end{equation}

This, and other expressions involving $\mu, \nu$ should be regarded as
essentially formal ones indicating the various properties of the
representation,
rather than to be used for computational purposes. For that, we work directly
in
the $x,p,\rho,\Pi$ representation.

We now come to the physical meaning of the extended Q-function: it is a
generalized probability distribution function defined on this semiquantal
phase space. To see this, we consider the properties of the
Q-function, which we can obtain as a slice of $Q_{ext}$.
In particular, the definition Eq. \ref{eq:Q} can be used to show that
\begin{equation}
\int d^2 \alpha \; Q(\alpha, \alpha ^*) = 1
\end{equation}
and that the averages of {\em antinormally } ordered products of the creation
and destruction operators is
\begin{equation}
<\hat a ^r (\hat  a ^{\dagger})^s> =\int d^2 \alpha \; \alpha ^r (\alpha ^*)^s
Q(\alpha, \alpha ^*) .
\end{equation}
These are properties similar to those of classical probability
distribution functions. There are many discussions \cite{hillery,gard,feng94}
on the merits of the Q representation versus those of others like
the Wigner function and the P-representation. Each has been critically
examined for its merits and demerits. We confine ourselves to the observation
that in most cases it can be easily shown that one can move between the various
representations with the help of well-defined transformations, and that the
representations have properties which render each convenient for different
calculations. We do, however, want to address the popular misconception that
the Q-function (or the Husimi function) is an inadequate representation in that
it cannot be used to reconstruct the original projection operator.
As is known in the quantum optics literature, $Q(\alpha, \alpha ^*)$ can be
expressed as an absolutely convergent power series in $\alpha$ and $\alpha ^*$.
To wit, since $Q(\alpha,\alpha ^*)$ can be written as
\begin{equation}
Q(\alpha,\alpha ^*) = e^{ (-\alpha \alpha ^*)} \sum_{n,m}{<n|\hat P|m>
\over \pi \sqrt{n!m!}} \alpha ^m (\alpha^*)^n,
\end{equation}
we can use the fact that for any projection operator $\hat P$, the
number state matrix elements satisfy
\begin{equation}\label{eq:nm}
|<n|\hat P|m>| \leq 1.
\end{equation}
to show that the double power series for $Q(\alpha, \alpha ^*)$ is
absolutely convergent (see Gardiner \cite{gard} for details).

If we write this power series as
\begin{equation}
Q (\alpha, \alpha ^*) = \sum_{n,m}Q_{n,m} \alpha ^m (\alpha^*)^n,
\end{equation}
the number state matrix elements of $\hat P$ can be explicitly written as
\begin{equation}
<n| \hat P|m> = \pi \sqrt{n!m!} \sum _r Q_{n-r,m-r}.
\end{equation}
Alternatively, $\hat P$ may be written as a {\em normally ordered} power series
\begin{equation}
\hat P = \pi \sum Q_{n,m}(\hat a ^{\dagger})^n \hat a ^m.
\end{equation}
Thus, the projection operator is obtained by replacing $\alpha ^m (\alpha^*)^n$
in the power series for $Q (\alpha, \alpha ^*)$ by the corresponding
{\em normally ordered} product of the operators $\hat a $ and
$\hat a ^{\dagger}$.

This discussion therefore provides us with a way of
constructing an extended Q-function for any wavefunction. We have also shown
how to reconstruct the corresponding density matrix (or projection operator)
from $Q_{ext}$: first project down onto
$Q (\alpha, \alpha ^*) $, and then use the power series expansion mentioned
above. It should be pointed out that this is a consequence of
the overcompleteness of Gaussians. A representation in a basis set with the
usual properties of orthonormality would not permit this reconstruction.

We also note that the eigenstates of $x$ and $p$ are contained
as asymptotic limits of the  squeezed coherent states,
hence the reduced position and momentum distribution
functions of the wavefunction $|\Psi>$ may be obtained directly from $Q_{ext}$.
To get this, we evaluate $Q_{ext}$ along $\mu = \delta \nu$ with $\delta$ real.
The limit $\mu \to \infty $ and $ \delta \to 1$ is $|\Psi (x)|^2$.
and  $\mu \to \infty,$ and $ \delta \to -1$ yields  $|\Psi (p)|^2$ (see Yuen
\cite{yuen} for details).
This is simpler in the $\rho, \Pi$ representation. The two limits are now
taken with $\Pi = 0$ and $\rho  \to 0$ and $ \rho \to \infty$ respectively.
In essence, since $Q_{ext}$ is defined on the space of all Gaussians, we
have taken advantage of the fact that the $\delta$-function can be obtained
as an infinitely squeezed Gaussian \cite{gas}.

The extended Q-representation is hence a valid representation. Moreover, the
the dynamics of arbitrary wavefunctions in this representation
are very simple. The von Neumann equation for the evolution of the density
matrix is replaced by a Liouville equation, by {\em exactly}
the same route that led to the Heisenberg equations of motion for arbitrary
operators being replaced by Poisson bracket equations:
\begin{equation}
{\partial  Q_{ext} \over \partial t} = \{H_{ext},Q_{ext}\}
\end{equation}
where $H_{ext}$ is, as before,
the expectation value of $\hat H$ taken with regard to the squeezed states.
The Poisson bracket is defined by Eq. \ref{eq:Poisson}.
Thus, we have constructed a semiquantal and a semiclassical propagator:
it is just the classical propagator for an arbitrary distribution function
\begin{equation}
U(t) = \exp (t\{H_{ext}, \}),
\end{equation}
the exponentiation of the Liouvillian for the extended Hamitonian system.
This propagator, by virtue of this construction, is well-behaved
and has nice properties - in particular it satisfies the composition property
\begin{equation}
U(t_1)U(t_2) = U(t_1 + t_2).
\end{equation}
This is the primary formal result of this section.
It is not quite a computational tool, however, since it corresponds
to the calculation of an infinite number of trajectories.

This discussion of phase space properties demonstrates
the formal power of this Hamiltonian approach: In particular,
the construction of a propagator with desirable properties.
In the following section, we show how the Hamiltonian
properties help us to obtain eigenfunctions and eigenvalues.

\section{Quantization}
We now sketch how to use these dynamics for requantization \cite{fn2} in
two different ways.
The first, which uses the Fourier Transform (FT) of the survival probabilty
$(S(t) \equiv <\Psi (0)|\Psi (t)>$) is a computational method in that
it is geared towards the numerical computation of eigenfunctions and
eigenvalues, directly using the dynamics of the wavefunction.
The second is a more formal consideration of the invariance properties of
periodic orbits and invariant torii, which enables the construction of a
quantization rule associated with these structures.
These separate considerations helps us bring out different
properties of these dynamics. In particular, they help understand the spectrum
obtained in terms of the phase space structures.

\subsection{Quantization using the survival probability}
As has been pointed out \cite{heller}, any approximate method
of computing the time-dependent behaviour of a wavepacket can be used to
construct a spectrum. The argument goes as follows.
Note that an expansion of the wavepacket (at any time $t$) in
terms of the eigenfunctions of the Hamiltonian exists, since they
form a complete set:
\begin{equation}
|\Psi (t)> = \sum c_n(t)|\phi _n>.
\end{equation}
All the time dependence is in the complex coefficients, since
the states $|\phi _n>$ are stationary states and the above can be written as
\begin{equation}
|\Psi (t)> = \sum c_n(0)e^{-iE_nt/\hbar}|\phi _n>.
\end{equation}
$E_n$ is the eigenvalue of the appropriate eigenfunction :
\begin{equation}
<\phi _n|\hat H |\phi _n> = E_n
\end{equation}
This implies that the survival probability can be written as
\begin{equation}
S(t) = <\Psi (0)|\Psi (t)> = \sum |c_n(0)|^2 e^{-iE_nt/\hbar}
\end{equation}
by orthonormality :
\begin{equation}
<\phi _n|\phi _m> = \delta _{nm}.
\end{equation}
It is now straightforward that the Fourier transform of
$S(t)$ yields the spectrum of the Hamiltonian.
Extracting the eigenfunctions is equally straightforward; the Fourier integral
of the wavefunction provides these:
\begin{equation}
\int_{-\infty}^{\infty}e^{iE_mt/\hbar}|\Psi (t)> = c_m|\phi _m>
\end{equation}
where we use the spectrum obtained from above to project out the $mth$ state.

The exact spectrum is available only when $|\Psi (t)>$ can be exactly computed.
Any approximation can be used to generate a spectrum, however, which is
in some sense as good (or as bad) as the approximate dynamics itself.
When we use any of the extended dynamics above for the case of the harmonic
oscillator, since these approximations are exact, it is easy to show that
this yields the exact spectrum and eigenfunctions \cite{heller,diss}.
For this case, the $(x,p)$ variables decouple from the $(\rho,\Pi)$ variables,
and all initial conditions give us periodic orbits (POs).
We take advantage of these POs to evaluate (in this case, analytically)
the infinite Fourier integrals above.
[Note that this is already better than the WKB  method, which does get the
spectrum right (with a little bit of work to evaluate the Maslov indices),
but does {\em not} yield the correct eigenfunctions.]
In general, for Heller's semiclassical dynamics, one has to work harder to get
wave-packet quantization, even if one does find POs. This is because these
POs are in the classical phase space and do {\em not} yield periodic behaviour
in the fluctation variables. This is true even when they are stable POs.

However for the Hamiltonian constructions, even for general systems,
if a PO can be found in the extended phase space (irrespective of stability),
the Fourier integrals for the spectrum and eigenfunctions can then be done
exactly.  This is a remarkable point:
in principle, a {\em single} periodic orbit can be used to extract the entire
spectrum of the system. There are some caveats, of course:
a) Most POs are found through symmetry considerations; they cannot then provide
the full spectrum and will only reflect the appropriate symmetry.
b) The weight factors $c_m$ for most eigenfunctions will be small, for a
general
PO. Hence, it will not be practically possible to extract these
eigenvalues and their eigenfunctions.  In practice, therefore, several POs
will be needed.
There has been a lot of effort in recent years to find POs in Hamiltonian
systems, directed primarily at Trace Formula work \cite{POs}.
This method is thus fortuitously situated.

The features of the spectra obtained through the survival probability method
(working with Heller's semiclassical dynamics) are not fully explained
however, \cite{Gutz}. A primary problem is that there is has not been, so far,
any means of verifying these calculations. The only comparisons one could make
is with spectra obtained through different methods (basis-set quantization, for
example). The discrepancies could then be put down to the differing nature of
the approximations. Working within the Hamiltonian formulations of wavepacket
dynamics, however, enables us to construct another quantization procedure:
a {\em different} approach within the {\em same} approximation.
This second method is what we might term invariance quantization.
The mathematical arguments are essentially the same as those for
Bohr-Sommerfeld
quantization and are detailed elsewhere \cite{diss,Kan}; we provide here a
simple constructive version.

\subsection{Berry's phase and quantization}
Consider the phase space trajectories obtained by solving the ordinary
differential equations (obtained above) for the parameters of the wavefunction.
We can see readily that a periodic orbit solution to Hamilton's equations
is {\em invariant} (on the ${\cal R}(2N)$ subspace) under the action of the
Hamiltonian. Thus, a wavefunction constructed as a sum over all the points
of the PO (each point corresponding to a Gaussian wave-packet) is a candidate
to be an approximate eigenfunction for the system. This ignores the role of
the phase factor: each point along the orbit acquires a
phase factor during the evolution. The dynamical phase is no problem, since it
is the same for all points along the orbit and can thus be factored as a global
phase. The geometrical part $\lambda _G$ for an arbitrary PO, however,
destroys the invariance in the full semiquantal space ${\cal R}(2N)
\times S(1)$. We must therefore consider the evolution of $\lambda _G$ as well.
We now note that a $PO \times \lambda _G$ such that the periodic evolution
of $\lambda _G$ on $S(1)$ is {\em commensurate} with that of the PO on
${\cal R}(2N)$ constitutes a function invariant on ${\cal R}(2N) \times
S(1)$. This, then, is our eigenfunction. The requirement of the commensurate
evolution of the phase translates simply to the relation
\begin{equation}
\lambda _G (PO) = \oint _{PO} {\bf P \cdot dQ} = n h ,
\end{equation}
where we have used Eq. \ref{eq:lg} and ${\bf P} = (p, \Pi) $ and
${\bf Q}= (x,\rho) $ as before.
This consideration of POs such that the point orbit acquires a Berry's phase
equal to integer multiples of Planck's constant $h$
has precisely the same form as the 'old' quantization rule of Bohr and
Sommerfeld. However, since the rule applies in the {\em extended} phase space,
as opposed to the classical phase space,
it does not have the same meaning. In particular, there are no
Maslov-Morse corrections \cite{EBK} to this rule, since there are
no singularities in the Gaussian representation.
It can be shown \cite{diss}, however, that the 'spread' variables
$\rho, \Pi$ explicitly take care of these corrections.
This had earlier been demonstrated through elegant group theoretical arguments
by Littlejohn \cite{lj} for the case of Heller's dynamics.
In general, the quantization condition above
will give us results different from the EBK rule (the POs are in the extended
space) but will always incorporate the Maslov correction.

We can thus see how to quantize using these POs in the Hamiltonian approaches:
the eigenfunction is constructed as a weighted sum over the
commensurate periodic orbit. The weight factor for each point along the orbit
is precisely the appropriate geometrical phase. The eigenvalue is $H_{ext}$ for
that orbit (or $H_{sc}$ if we are using the truncated form).
It is interesting to note that this quantization condition arises
{\em entirely} from a consideration of Berry's phase for the dynamics.

The extension of this argument from POs to invariant torii goes through
easily \cite{diss,Kan} and leads to a general quantization rule:
\begin{equation}\label{eq:pdq}
\oint _{C_i} {\bf P \cdot dQ} =  n_i h
\end{equation}
where the closed integral is now taken over the $ith$ irreducible contour
around
the torus and the quantum numbers $n_i$ are labeled accordingly. This is
exactly Einstein's generalization \cite{Einstein} of the Bohr-Sommerfeld rule
to invariant torii.
It has been shown \cite{Kan,Lj,diss} that the eigenfunctions (or rather their
Q-functions) are, by the construction above, peaked around these invariant
torii \cite{scars}.

We emphasize that this method can be used in either of the Hamiltonian
formulations constructed above. We now outline an example ; the details are the
subject of a forthcoming paper \cite{diss}.

\subsection{Quantization of a chaotic Hamiltonian}

As an example of the methods formulated above, we have studied the Hamiltonian
\begin{equation}
\hat H = {\hat p_x^2 + \hat p_y^2 \over 2} + {1 \over 2}\hat x^2 \hat y^2
+ \beta(\hat x^4 +\hat y^4).
\end{equation}
This Hamiltonian has recently been the subject of some large-scale numerical
work \cite{Eckhardt}.
These indicate that the classical version ($\hat O \to O$ for all
operators) is a very strongly chaotic system, with few, if any, stable periodic
orbits in the limit $\beta \to 0$. The quantal Hamiltonian with $\beta =0$
also resists numerical analysis: basis-set quantization with matrices of
dimension $3240$ do not provide converging eigenvalues (see the paper by
Eckhardt {\em et al} \cite{Eckhardt} for details). They have hence used $\beta
=0.01$ for their analysis. As they point out, the eigenfunctions of
this Hamiltonian belong to the symmetry classes of of the $C_{4\nu}$ symmetry
group which has eight elements ($4$ reflections in the axes and diagonals and
$4$ rotations by $\pi/2$). The irreducible representations of this group split
into four one-dimensional representation and one two-dimensional
representation.
They have restricted themselves to the four one-dimensional representations and
used harmonic oscillator basis sets to obtain low-lying eigenvalues and
eigenstates for this system.
We have applied the semiquantal method detailed above to this system.
It works very well; in particular it does excellently  for
$\beta=0$.  The extended Hamiltonian in that case is
\begin{eqnarray}
H_{ext} & =&  {1 \over 2}(p_x^2 +p_y^2 + \Pi_x^2 + \Pi_y^2)\nonumber\\
 & + &   {1 \over 8 \rho _x^2} + {1 \over 8 \rho _y^2}
+ { 1 \over 2}(x^2 + \rho _x^2)(y^2 + \rho _y^2).
\end{eqnarray}

If we consider the same one-dimensional representations as Eckhardt {\em et al}
corresponding to wavefunctions which are A) symmetric under $x \to y, x \to
-x$,
B) antisymmetric, symmetric , C) symmetric, antisymmetric  and
D) antisymmetric, antisymmetric respectively, we find that all four subspaces
can be studied by the symmetry reduced version of $H_{ext}$ above:
\begin{equation}
H = { 1\over 2}(p^2 + \Pi ^2) + {1 \over 8 \rho ^2} +
{1 \over 4}(z^2 + \rho ^2)^2
\end{equation}
where $(z,p)$ and $(\rho, \Pi)$ are the canonically conjugate pairs.
This new Hamiltonian is explicitly integrable: it has two isolating integrals
of motion, the total energy, and an invariant obtained by construction,
\begin{equation}
I = ( \rho p - z\Pi)^2 + { z^2 \over 4\rho^2}.
\end{equation}
This fact enables us to find the invariant torii needed for the invariance
quantization detailed above, and evaluate the eigenvalues in a straightforward
fashion.
The details are provided in Ref. \cite{diss}; however, we note here that the
ground state energy we obtain analytically is $E_0 = 0.5953$.
This is substantially lower than the value obtained by Eckhardt {\em et al}
of $E_0^{num} = 1.093$. The semiquantal method is based on a time-dependent
variational principle; it is easy to show that for the ground state it
reduces to the time-independent (Rayleigh-Ritz) variational principle.
The semiquantal result thus provides an upper bound to the exact result and
hence improves upon the numerical results.
This is true even when we use $\beta =0.01$ in the semiquantal method. The
ground state energy in that case is $E_0^{\beta} =0.6012$,
only marginally different from the $\beta=0$ case, and still better than
the numerical result.

This example shows that the semiquantal method can be used to advantage
to obtain eigenvalues and eigenfunctions for classically chaotic systems and
at least in some cases give excellent results.
It also tells us what role of semiquantal dynamics is in the
agenda of post-modern quantum mechanics: it is
the search for a variational space such that the dynamics on this
space are explicitly integrable. We then apply the rule \ref{eq:pdq} to
quantize the system, hence fulfilling the program of the extension of the
'old' quantization rule, and advancing the comprehension of the
quantum-classical correspondence limit.

\section{Summary}
In this paper, we have developed Gaussian semiquantal dynamics to provide a
phase space formalism. We have thus been able to construct a propagator with
desirable qualities. We have compared these equations to the standard
semiclassical Gaussian dynamics and have shown that the latter
arise as truncations to the semiquantal equations {\em not self-consistent} in
$\hbar$, though consistent in derivatives of the Hamiltonian.
This has enabled us to
extend the usual treatment and introduce a Hamiltonian semiclassical dynamics
that retains the power of the phase space formulation.
We have then qualitatively evaluated the behaviour of these three different
approaches in various situations and have shown that the semiquantal dynamics
substantially improve upon the standard semiclassical equations in
reproducing the expected quantal behavior and the extended semiclassical
dynamics do so marginally.
Finally, we have shown
how to obtain approximate eigenvalues and eigenfunctions in this formalism, and
have demonstrated with the help of an example that it works well even for a
classically chaotic Hamiltonian.
\acknowledgements
A.K.P. wishes to acknowledge partial support from the Robert A. Welch
Foundation (Grant No. F-0365).
\appendix
\section*{Generalization to N dimensions}

In this appendix, we discuss the generalization of the Hamiltonian formulation
of semiquantal and semiclassical dynamics to $N$ dimensions.
The simplest way of doing this is to use product spaces for the variational
subspace, for example,
\begin{equation}
\Psi (x,y) = \Psi _x (x) \Psi _y (y).
\end{equation}
With this factorization, all the one-dimensional arguments carry through
exactly to $N$ dimensions, and an extended potential formulation (with a
semiquantal phase space of $4N$) can be shown to exist \cite{diss}.

If we allow for correlations between the various degrees of freedom, however,
this is no longer possible. The Hamiltonian formulation still exists
\cite{KraSa}, but the extended potential cannot be extracted.
The best way to derive the equations of motion in this case is to use
the TDVP and a wavepacket $< {\bf x} |\Psi, t>$ of the form
\begin{equation}
\Psi ({\bf x}, t) = \exp\bigg [{ i \over \hbar }[ {\bf (x - q)\cdot A \cdot
(x - q)   +  p\cdot(x-q) }]\bigg ].
\end{equation}
where now ${\bf x, p, q}$ are vectors and ${\bf A}$ a symmetric $N \times N$
matrix.
If we continue to assume a classical Hamiltonian of the form
\begin{equation}
H_{classical} = {1 \over 2} {\bf p ^2} + V({\bf x})
\end{equation}
and rewrite ${\bf A}$ as
\begin{equation}
{\bf A} = i{1 \over 4} {\bf G^{-1}} + {\bf K}
\end{equation}
with ${\bf G, K}$ real symmetric matrices, we can obtain the extended
Hamiltonian
\begin{eqnarray}
H_{ext} & =& H_{classical} + Tr( {1 \over 8}{\bf G^{-1} + K^2G + GK^2})
\nonumber\\
&+ &<V({\bf \hat x})>.
\end{eqnarray}
where $Tr$ indicates the trace operation on the matrices.
The only calculation of any subtlety is the evaluation of $<V({\bf \hat x})>$,
since $\Psi ({\bf x})$ is a multivariate Gaussian distribution. The result is a
series, as for the one-dimensional case, in even derivatives of $V$.
The coefficients of these terms are found by the standard prescription:
\begin{equation}
<X_iX_jX_k\cdots> = \sum <X_aX_b><X_cX_d> \cdots
\end{equation}
where the subscripts $a,b,c,d,\ldots$ are the same as $i,j,k\ldots$ taken two
by two. The summation extends over all the ways in which $i,j,k\ldots$ can be
subdivided in pairs. The terms on the right-hand-side of this equation are
just the ${N^2 + N \over 2}$ elements of ${\bf G}$, which
is the covariance matrix of the multivariate distribution. These elements
constitute the canonical coordinates (along with the vector ${\bf x}$ ) of
the extended Hamiltonian; the corresponding canonical momenta are the elements
of ${\bf K}$ (and the vector ${\bf p}$).
If the two matrices are chosen diagonal, the system reduces to the product
space discussed above, and the extended potential exists. It is the presence
of the cross terms in ${\bf K^2G + GK^2}$  that prevents this in the general
case. All arguments of the Hamiltonian formulation, including those about
truncations and reduction to semiclassical dynamics carry through,
irrespective: there are a few more equations, and they do not have as simple a
form, however.
We also note that the global phase of the wavefunction is, exactly as before,
the extended action evaluated along the trajectory in the extended space.


\begin{thebibliography}{99}
\bibitem{phystoday} E.J. Heller and S. Tomsovic, { Physics Today}, July 1993,
pg. 38.

\bibitem{stev}P. Stevenson, { Phys. Rev.} {\bf D30}, 1712 (1984);
{\it ibid} {\bf 32}, 1389 (1985) and references therein. This discussion of
effective potentials was applied to chaos in quantum systems in
L. Carlson and W.C. Schieve, {\sl Phys. Rev. A} {\bf 40}, 5896 (1989) and
A.K. Pattanayak and W.C. Schieve, {\sl Phys. Rev. A} {\bf 46}, 1821 (1992);

\bibitem{prl}A.K. Pattanayak and W.C. Schieve, {Phys. Rev. Lett}
{\bf 72}, 2855 (1994) and references therein.

\bibitem{EBK}I.C. Percival, {Adv. Chem. Phys.} {\bf 36}, 1 (1977)
provides a review of this.

\bibitem{gutz} M. Gutzwiller, {\it Chaos in Classical and Quantum Mechanics},
(Wiley-Interscience, 1991).

\bibitem{linda} L.E. Reichl, {\it The transition to chaos :Quantum
manifestations}, (Springer-Verlag, 1992) and references therein.

\bibitem{eckhardt} B. Eckhardt, {Phys. Rep.} {\bf 163}, 205 (1988).

\bibitem{leshouches} M.-J. Giannoni, A. Voros and J. Zinn-Justin (Editors),
{\it Chaos and Quantum Physics}, Proceedings of the Les Houches Summer School
1989, (North-Holland, 1991).

\bibitem{oconnor} P.W. O'Connor,J. Gehlen and E.J. Heller {Phys. Rev.
Lett.} {\bf 58}, 1296 (1987).

\bibitem{meredith} D.C. Meredith, {J. Stat. Phys.} {\bf 68}, 97 (1992)
provides a recent review of this.

\bibitem{jensen} R.V. Jensen, M.M. Sanders, M. Saraceno and B. Sundaram,
{Phys. Rev. Lett.} {\bf 63} 2771 (1989).

\bibitem{delos} J.B. Delos and R.T. Swimm,{Chem. Phys. Lett.} {\bf 47}, 76
(1977) and references therein.

\bibitem{hillery} See  M. Hillery, R.F. O'Connell, M.O. Scully and E.P. Wigner,
{Phys. Rep.} {\bf 106}, 121 (1984) for a review.

\bibitem{feng94} W.-M. Zhang and D.H. Feng , 'Quantum Nonintegrability',
 {Phys. Rep.} (In press, 1994); Mod. Phys. Lett {\bf A8}, 1417 (1993).

\bibitem{klauder} J.R. Klauder, {J. Math. Phys.},{\bf 4}, 1055 (1963);
1058 (1963).

\bibitem{sudarshan} E.C.G. Sudarshan, {Phys. Rev Lett.} {\bf 10}, 277 (1963)

\bibitem{glauber} R. Glauber,{Phys. Rev. Lett.}{\bf 10}, 84 (1963);
{Phys. Rev.} {\bf 130}, 2529 (1963).

\bibitem{witten} E. Witten, {Nucl. Phys.} {\bf B160}, 57 (1979).

\bibitem{yaffe} L.G. Yaffe, {Rev. Mod. Phys. }{\bf 54}, 407 (1982).

\bibitem{skag}J. Klauder and B.-S. Skagerstam (Editors.), {\it Coherent States:
Applications in Physics and Mathematical Physics} (World Scientific, 1985)
is a fine review.

\bibitem{rmp}  W.-M. Zhang, D.H. Feng, and R. Gilmore, {Rev. Mod. Phys.},
{\bf 62}, 867 (1990).

\bibitem{perel} A.M. Perelmov, {\it Generalized coherent states and their
applications} (Springer-Verlag, 1986).

\bibitem{heller} E.J. Heller, {J. Chem. Phys.} {\bf 62}, 1544 (1975); his
Lecture Notes in Ref. \cite{leshouches} is an excellent summary of his work.

\bibitem{Schro}Some of these ideas date back to E. Schr\"oedinger,
{Naturwissenschaften}, {\bf 4}, 664 (1926).

\bibitem{LJ} R.G. Littlejohn, {Phys. Rep.} {\bf 138}, 193 (1986).

\bibitem{Jackiw} R. Jackiw and A. Kerman, Phys. Lett. {\bf 71A}, 158 (1979).

\bibitem{cooper} F. Cooper, S.-Y. Pi and P.N. Stancioff,
{Phys. Rev.} {\bf D34}, 3831 (1986).

\bibitem{kovner} A. Kovner and B. Rosenstein, {Phys. Rev.} {\bf D39}, 2332,
(1989).

\bibitem{dirac} P.A.M. Dirac, Appendix to the Russian edition of {\it The
principles of quantum mechanics}, as cited by I.I. Frenkel, {\it Wave
mechanics,
advanced general theory} (Clarendon Press, Oxford, 1934) (pgs. 253, 436).
We apologise for being unable to trace a reference other than this rather
esoteric one; the mystery is deepened by the fact that the standard citation of
P.A.M. Dirac, {Proc. Cam. Phil. Soc.} {\bf 26}, 376 (1930) does not contain
this equation.

\bibitem{KraSa} P. Kramer and M. Saraceno, {\it Geometry of the Time-Dependent
Variational Principle in Quantum Mechanics}, (Springer-Verlag, 1981).

\bibitem{diss} A.K. Pattanayak and W.C. Schieve, in preparation;
A.K. Pattanayak, Ph. D. dissertation, The University of Texas
at Austin, 1994 and references therein.

\bibitem{stoler} D. Stoler, { Phys. Rev.} {\bf D1}, 3217 (1970);{\it ibid}
{\bf D4}, 1925 (1971); D. F. Walls, {Nature} {\bf 306}, 225 (1983).

\bibitem{andrews} For other finite truncations, see also M. Andrews,
{J. Phys.} {\bf A14}, 1123 (1981) and Ref. \cite{diss}.

\bibitem{tsue} Y. Tsue, {Prog. Theor. Phys. } {\bf 88}, 911 (1992)
and references therein.

\bibitem{Rajagopal}A slightly different version of these equations were also
obtained by
A.K. Rajagopal and J.T. Marshall, {Phys. Rev.} {\bf A27} 558, (1982).
We thank Prof. Rajagopal for this reference.

\bibitem{wilczek}This is actually the Aharanov-Anandan version of Berry's
phase. See the reprint collection {\it Geometric Phases in Physics}, Eds.
A. Shapere and F. Wilczek, (World Scientific, 1989) for details

\bibitem{Voros} J. Kurchan, P. Leboeuf and M. Saraceno, {Phys. Rev.}
{\bf A40}, 6800 (1989); A. Voros, {Phys. Rev.} {\bf A40}, 6814 (1989).

\bibitem{hage} G.A.Hagedorn, {Commun. Math. Phys.} {\bf 71} 77 (1980).

\bibitem{gard} C.W. Gardiner, {\it Quantum Noise}, (Springer-Verlag,1990).

\bibitem{yuen}H.P. Yuen, {Phys. Rev.} {\bf A13}, 2227 (1976).

\bibitem{fn1}Actually, since we have argued above that any {\em particular}
choice of $\mu, \nu$ such that Eq.\ref{eq:munu} is satisfied leads to an
equivalent set of states, {\em any} particular slice of $Q_{ext}$ yields an
Q-function equivalent to $Q(\alpha, \alpha ^*)$.

\bibitem{gas} S. Gasiorowicz, {\it Quantum Mechanics}, pg. 493 (John Wiley
\& Sons) (1974).

\bibitem{fn2} We use this term to reflect the idea that we had reduced the
quantum system to a classical problem different from the canonical one. We are
now {\em returning} to the quantum notions of eigenfunctions and eigenvalues.

\bibitem{POs} See the issue dedicated to periodic orbit theory, Chaos {\bf 2},
(1992).

\bibitem{Gutz} M. Gutzwiller, {\it op. cit}, pg. 238.

\bibitem{Kan} K.K. Kan, {Phys. Rev.} {\bf C24}, 279 (1981); E. Caurier,
S. Drozdz and M. Ploszajczak, {Phys. Lett.} {\bf A134}, 1 (1984) and
references therein.

\bibitem{lj} R.G. Littlejohn, {Phys. Rev. Lett.} {\bf 61}, 2159 (1988).

\bibitem{Einstein} A. Einstein, {Verh. Dtsch. Phys. Ges.} {\bf 19},
82 (1917); transalated by C. Jaffe, Report 116, JILA, Boulder, CO (1980).

\bibitem{Lj} R.G. Littlejohn, {Phys. Rev. Lett.} {\bf 56}, 2000 (1986).

\bibitem{scars}  While these are not scars, which are associated
with the classical phase space, they are their semiquantal analog.
These have then to be projected from the extended phase space
onto the classical phase space to compare with scars.

\bibitem{Eckhardt} B. Eckhardt, G. Hose and E. Pollack, {Phys. Rev.} {\bf A39},
3776 (1989) and references therein.

\end{thebibliography}
\end{document}